\documentclass[twocolumn,showpacs,preprintnumbers,amsmath,amssymb,prb]{revtex4}

\usepackage{graphicx}
\usepackage{dcolumn}
\usepackage{bm}
\usepackage{wasysym}
\usepackage{subfigure}
\usepackage{float}
\usepackage{wrapfig}

\begin{document}

\preprint{}

\title{Gap Opening by Asymmetric Doping in Graphene Bilayers}

\author{Marcos G. Menezes}
 \email{marcosgm@if.ufrj.br}
\author{Rodrigo B. Capaz}%
 \email{capaz@if.ufrj.br}
\affiliation{Instituto de F\'{i}sica, Universidade Federal do Rio de Janeiro, Caixa Postal 68528 21941-972, Rio de Janeiro, RJ, Brazil}%

\author{Jorge L. B. Faria}
 \email{hulk@fisica.ufmt.br}
\affiliation{Instituto de F\'{i}sica, Universidade Federal do Mato Grosso, 78060-900, Cuiab\'a, MT, Brazil}%

\date{\today}

\begin{abstract}
We study the energy gap opening in the electronic spectrum of graphene bilayers caused by asymmetric doping. Both
substitutional impurities (boron acceptors and nitrogen donors) and adsorbed potassium donors are considered. The gap evolution with dopant concentration is compared to the situation in which the asymmetry between the layers is induced by an external electric field. The effects of adsorbed potassium are similar to that of an electric field, but substitutional impurities behave quite differently, showing smaller band gaps and a large sensitivity to disorder and sublattice occupation.
\end{abstract}

\pacs{71.15.Ap, 71.15.Dx, 71.15.Mb, 71.15.Nc}
\maketitle

\section{\label{intro}Introduction}
Graphene materials show unusual structural and electronic properties, attracting the attention of the scientific community for its potential applications in nanoeletronics.\cite{rmp} However, to realize this potential, it is important to make graphene semiconducting. Therefore, bilayer graphene (BLG), a double layer of graphene with AB stacking, is very attractive, since it is a metallic system but it can be made semiconducting by breaking the symmetry between the two layers \cite{mccann-falko06,guinea06,mccann06,macdonald,neto,ohta,zhou08,oostinga08,mak09,zhang09}. More interestingly, the gap can be tuned by the magnitude
of the symmetry breaking, which can be experimentally produced by doping \cite{ohta,zhou08} or by an external and perpendicular electric field \cite{oostinga08,mak09,zhang09}. 

So far, virtually all theoretical calculations of this effect have considered the case of a spatially homogeneous  symmetry breaking (with respect to in-plane coordinates), which can be experimentally realized by an external electric field. However, it is not clear if this approach can describe equally well the symmetry breaking induced by doping, in which the effects of spatial inhomogeneity, disorder and chemical bonding are present. 

In this work, we address precisely these issues, by a combination of {\it ab initio} density-functional theory (DFT) and tight-binding (TB) calculations. We consider three specific situations: boron (acceptor) or nitrogen (donor) as substitutional impurities and potassium adsorption (donor) in one of the sheets. By comparing our results with the case of a homogeneous electric field, we are able to understand and to predict the situations in which the spatially homogeneous modeling provides a good description of the gap opening induced by impurity doping. Furthermore, for the substitutional case, TB calculations in large supercells are used to investigate the influence of sublattice disorder in the electronic gap.

Our article is divided as follows: We begin by detailing our calculation method in the next Section. Then, in Section \ref{band} we present our band structure calculations and compare the actual doping cases with the homogeneous field model for several ordered configurations of impurities. Next, in Section \ref{disorder} we study effects of disorder using large supercells within the TB model. Finally, in Section \ref{conclusion}, our conclusions are presented.

\section{\label{met}Methodology}
\begin{figure}[h]
\centering
\includegraphics[width=10cm]{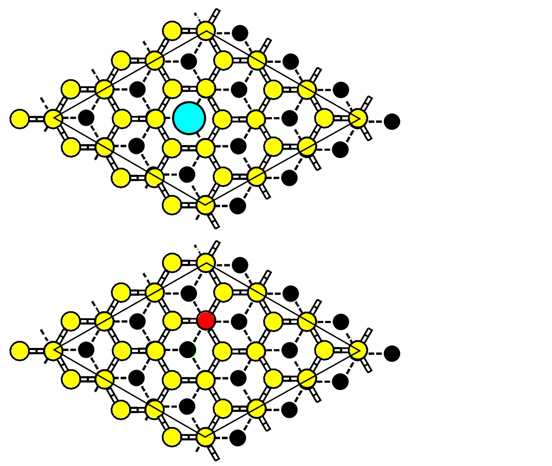}
\caption{\label{fig-cells} Some representative $3\times 3$ supercells studied in this work. Upper panel: Potassium adsorption. Lower panel: Boron or nitrogen substitutional doping. Black, yellow, light blue and red represent carbon (lower sheet), carbon (upper sheet), potassium and boron or nitrogen atoms, respectively.}
\end{figure}

Our \textit{ab initio} calculations are based on the density functional theory \cite{hk64,ks65}, pseudopotential method, plane-wave basis and supercells with periodic boundary conditions, as implemented in the the PWSCF package.\cite{pwscf} We work with graphene bilayers with Bernal stacking and approximate 10 {\AA } of vacuum in the direction perpendicular to the sheets. The average distance between sheets is set to the experimental value for graphite (3.35 \AA). All other geometrical parameters are found by energy minimization until forces are smaller than $10^{-3}$ Ha/Bohr. Dipole corrections are employed in order to impose a zero electric field in the vacuum region. Hexagonal supercells of different lateral sizes $a$ are used in order to vary the dopant concentrations. In this work, we consider $2\times 2$, $3\times 3$, $4\times 4$, $5\times 5$ and $6\times 6$ supercells. Figure ~\ref{fig-cells} shows top views of representative systems for $a=3a_0$ ($3\times 3$) supercells, where $a_0 = 2.46$ {\AA } is the theoretical lattice constant for BLG. Substitutional (boron or nitrogen) and potassium adsorption are shown in Fig. ~\ref{fig-cells} (bottom and top panels, respectively). Nitrogen and boron substitutions occur in only one of the sheets, thus breaking the symmetry between the layers. We choose to position the substitutional impurities right on top of a carbon atom at the other layer. This choice does not affect our main results, as we learn from tests performed in other configurations. The Brillouin zone is sampled at $N_k \times N_k \times 1$ Monkhorst-Pack meshes\cite{mp}, in which $N_k$ is inversely proportional to the lateral size $a$, $N_k = C/a$, with $C \approx 12a_0$.
For substitutional impurities, a plane wave cutoff $E_{cut}=25$Ry is used, along with the generalized gradient approximation (GGA) and Vanderbilt ultrasoft pseudopotentials.\cite{vanderbilt} For the K adsorption case, a plane wave cutoff of $E_{cut}=40$Ry is used, along the local density approximation (LDA) and Von Barth Car (direct fit) pseudopotentials with non-linear core correction.\cite{dalcorso93} K atoms are placed above the hexagon centers of one of the layers.

\begin{figure}[h]
\centering
\includegraphics[width=10cm]{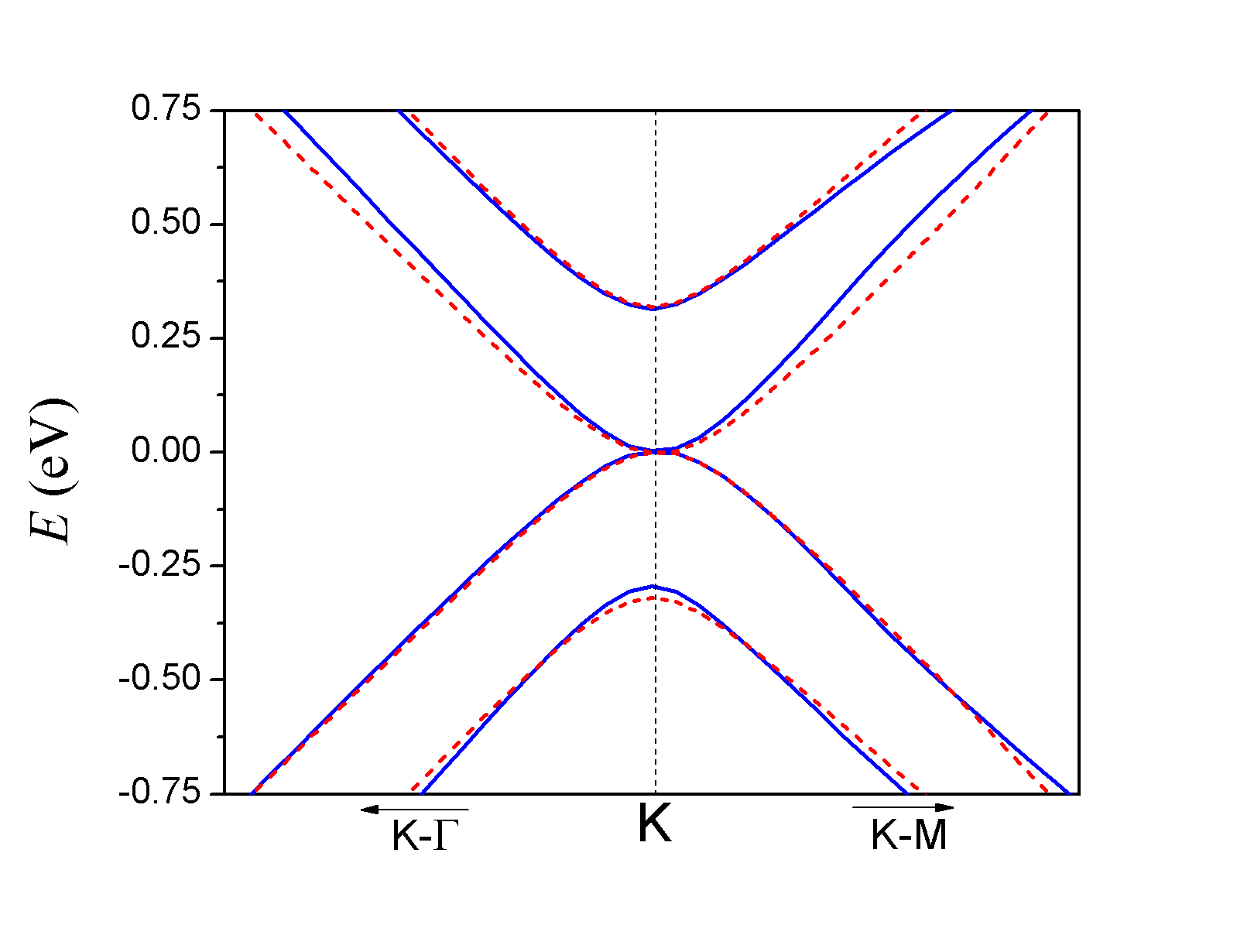}
\caption{\label{fig-bands-pristine} DFT (blue full line) and tight-binding (red dashed line) band structures of pristine BLG.}
\end{figure}

For the tight-binding calculations of substitutional impurities, we use the simplest basis set of one $p_z$ orbital per carbon atom within the Slonczewsky-Weiss-McClure (SWM) parametrization \cite{mcclure57,slonc58}, with in-plane hopping $\gamma_0 = -2.4$ eV, and interlayer hoppings $\gamma_1 = -0.32 $ eV ($A-A$ coupling) and $\gamma_3 = -0.30 $ eV ($A-B$ coupling). The onsite matrix elements for carbon atoms is chosen to be 0.0 eV. These parameters are chosen so as to reproduce the DFT band structure of BLG, as shown in Figure ~\ref{fig-bands-pristine}. TB calculations for BLG under a homogeneous electric field were performed by adding an energy $U$ to the onsite matrix elements of all the atoms in one of the layers, as described in detail in the next Section. Finally, a good description of the electronic structure of substitutional impurities can only be achieved by combining the introduction of onsite matrix elements at the impurity sites of $\epsilon_B = 4.2$ eV
for boron (the case of nitrogen is completely analogous and it was not considered) {\it and} a 
homogeneous-field-like $U$ term, as described in Section \ref{disorder}. 

\section{\label{band}Band Structure Results}
Initially, we focus our analysis on the similarities and differences between the band structures of BLG in two situations: (a) under a homogeneous and perpendicular external electric field and (b) under actual doping by donor and acceptor species. The DFT band structure of BLG under external electric field has been studied by Min {\it et al.} \cite{macdonald} and their results show that the simple tight-binding model described in the previous Section (including $\gamma_3$ hopping) describes extremely well the evolution of the energy gap as a function of
the difference in screened electrostatic potential between the two layers. Within the tight-binding description, this potential difference is described by adding an energy $U$ to the onsite matrix elements of all carbon atoms in one of the layers, while keeping all other parameters fixed. Morever, this parameter $U$ can be easily determined from the band structure, since it is precisely equal to the energy gap at the $K$ point of the Brillouin zone. Therefore, for simplicity, we adopt this tight-binding description for the case of a homogenous eletric field, but we emphasize once more that it provides results that are virtually identical to the DFT calculations.

\begin{figure}[h]
\centering
\includegraphics[width=10cm]{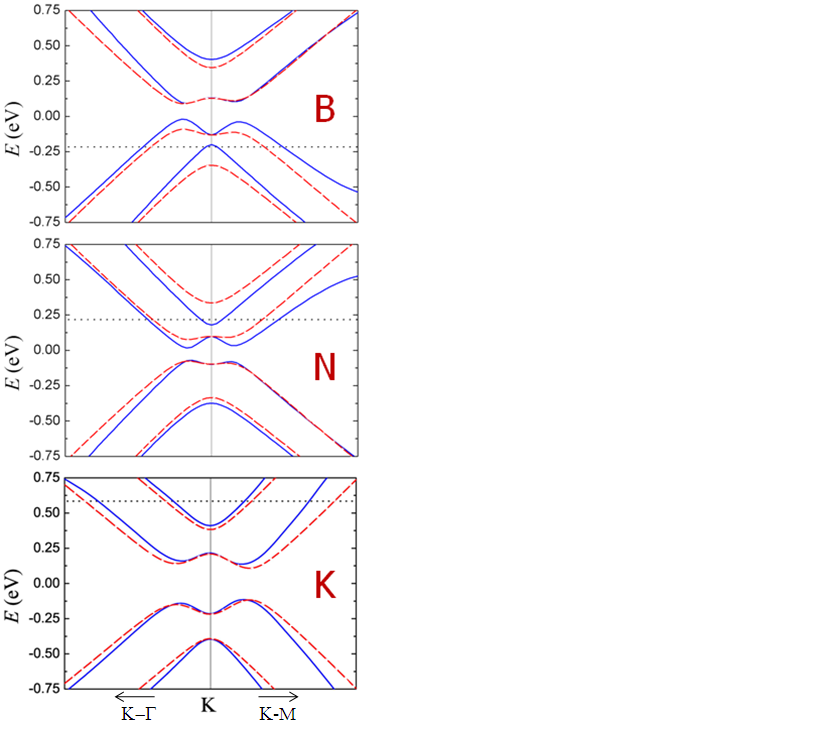}
\caption{\label{fig-4x4} Chemical doping (blue full line) and homogeneous electric field (red dashed line) band structures of doped BLG in a $4\times 4$ supercell. From top to bottom: boron, nitrogen and potassium doping. The dashed horizontal line indicates the Fermi level.}
\end{figure}

It is sufficient to analyse in detail the cases of $3\times 3$ and $4\times 4$ supercells, since they are representative of the other cases.  Figure ~\ref{fig-4x4} shows the band structure near the Fermi level for different dopants in a $4\times 4$ supercell. In this case the $K$ point of the BLG Brillouin zone is folded into the $K$ point of the supercell. The blue line corresponds to the actual doping situation while the red line is the homogeneous field case, in which $U$ is adjusted to reproduce the gap at the $K$ point. One notices that, for B and N substitutional doping, the homogeneous field description substantially underestimate the band gaps and the doping-induced particle-hole asymmetry is not reproduced. The agreement is better for potassium adsorption, but there are still noticeable differences in the band structure. 
 
\begin{figure}[h]
\centering
\includegraphics[width=10cm]{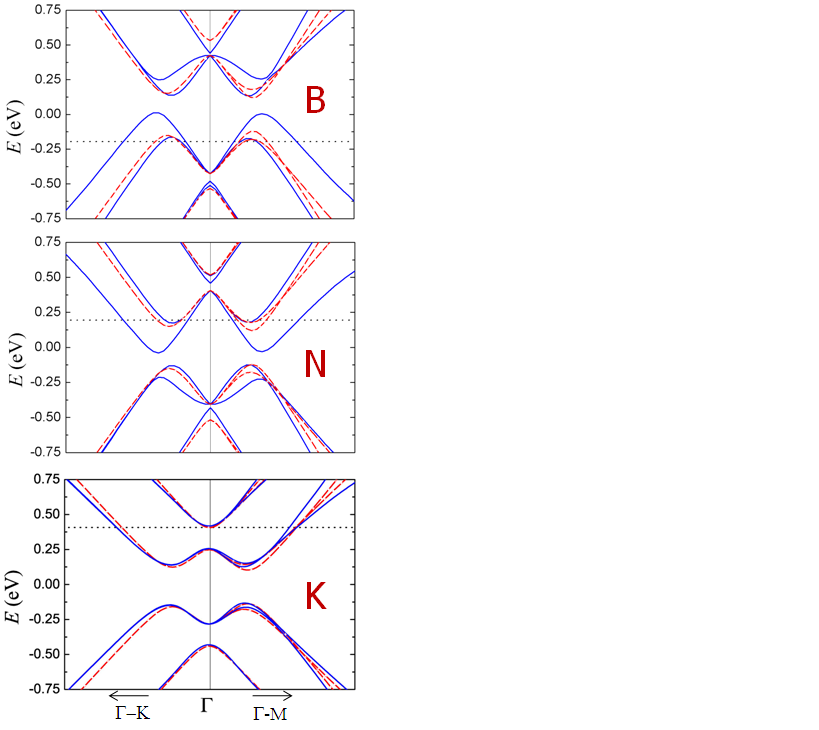}
\caption{\label{fig-3x3} Chemical doping (blue full line) and homogeneous electric field (red dashed line) band structures of doped BLG in a $3\times 3$ supercell. From top to bottom: boron, nitrogen and potassium doping. The dashed horizontal line indicates the Fermi level.}
\end{figure}
 
Figure ~\ref{fig-3x3} shows the band structure near the Fermi level for different dopants in a $3\times 3$ supercell. Again, the blue line corresponds to the actual doping situation while the red line is the homogeneous field case. In this particular supercell, both the $K$ and $K'$ points of the BLG Brillouin zone are folded into the $\Gamma$ point of the supercell, so Figure ~\ref{fig-3x3} displays twice as many bands as in the $4 \times 4$ case. More importantly, such folding produces degeneracy lifting and substantial level repulsion for the case of B and N doping. These effects are absent in the homogeneous field approximation, thus making much worse the agreement between this approximation and the actual doping case. In fact, one notices that gaps are indirect in DFT, and this effect is not captured by the homogeneous field approximation. Interestingly, the agreement is better for potassium adsorption. The reason is related to symmetry: contrary to adsorbed potassium atoms, that sit on top of hexagons,  B and N substitutional defects break the A/B sublattice symmetry, thus coupling more strongly $K$ and $K’$ points. Therefore, the homogeneous field approximation produces a fairly good description of the potassium adsorption case.

\begin{figure}[h]
\centering
\includegraphics[width=10cm]{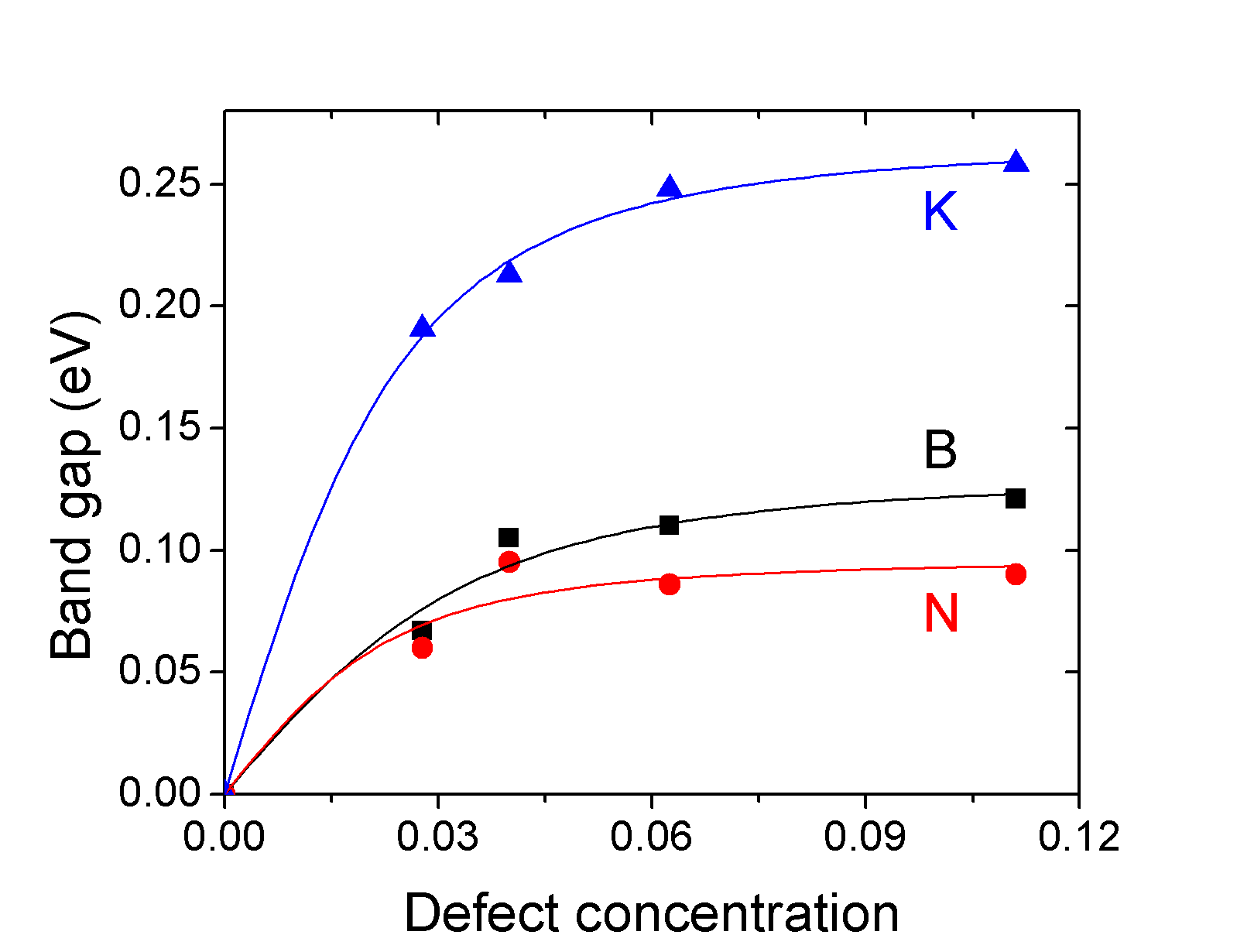}
\caption{\label{concentration} Band gap as a function of defect concentration for boron, nitrogen and potassium doping. Lines are guides to the empirical function described in the text.}
\end{figure}

For device applications, it is important to describe the band gap dependence as a function of defect concentration $\sigma=(a_0/a)^2$. This is analogous, in the case of a homogeneous electric field, to the gap dependence with respect to the field magnitude. The latter case has been extensively studied theoretically and it is well known that the band gap increases linearly with the field for low field values and then saturates at $E_{max} \approx 0.25$ eV. In fact, from the tight-binding model under the approximation of $\gamma_3 = 0$, we expect $E_g \approx U E_{max}/\sqrt{E_{max}^2 + U^2}$. As an {\it ansatz}, we expect that an equivalent or average $U$ for the doping case should be proportional to the defect concentration, $U = \alpha \sigma$. Therefore we propose the following empirical equation to describe the energy gaps as a function of defect concentration:
\begin{eqnarray}
E_g=\frac{\alpha \sigma E_{max}}{\sqrt{E_{max}^2 + (\alpha \sigma)^2}}.
\label{eq:conc}
\end{eqnarray}
Figure ~\ref{concentration} shows the minimum Kohn-Sham band gap as a function of defect concentration for all three dopants. Qualitatively, the doped systems behave similarly to case of a homogenous electric field and follow roughly Eq. \ref{eq:conc}, showing gap increase and saturation as the concentration increases. However, there are clear and important differences. First, notice that the gap saturates at much lower values (around 0.1 eV) for the substitutional dopants boron and nitrogen with respect to potassium doping (which in fact follows closely the homogeneous field prediction). Second, there are sizeable gap fluctuations around the empirical curves for the boron and nitrogen cases, probably reflecting the gap sensitivity to the particular ordered arrangement of impurities considered: most likely these fluctuations are averaged out if a disordered arrangement of impurities is considered. These effects cannot be captured by a homogeneous field description.

So far, our results show that the homogeneous field approximation provides a reasonable description of the gap opening in BLG by potassium adsorbates, but it seriously fails       
to describe the behavior of substitutional boron acceptors and nitrogen donors. Therefore, it is reasonable to question if the gap opening in these substitutional systems is robust to a disordered arrangement of impurities, in which both A and B sublattices of one of the layers are randomly occupied by the defects. We address this issue in the following Section.

\section{\label{disorder}Disordered arrangement of substitutional impurities}

In order to consider a disordered configuration of substitutional impurities, one needs to perform
calculations with very large supercells and configurational averages among different
realizations. For this purpose, the {\it ab-initio} approach becomes computationally prohibitive, therefore we adopt a tight-binding model. However, our tight-binding model must be able to reproduce the DFT band structure at least for an ordered configuration of substitutional impurities. This is clearly not the case for a simple tight-binding model with an homogenous electric field, as shown in Section III.

\begin{figure}[h]
\centering
\includegraphics[width=8cm]{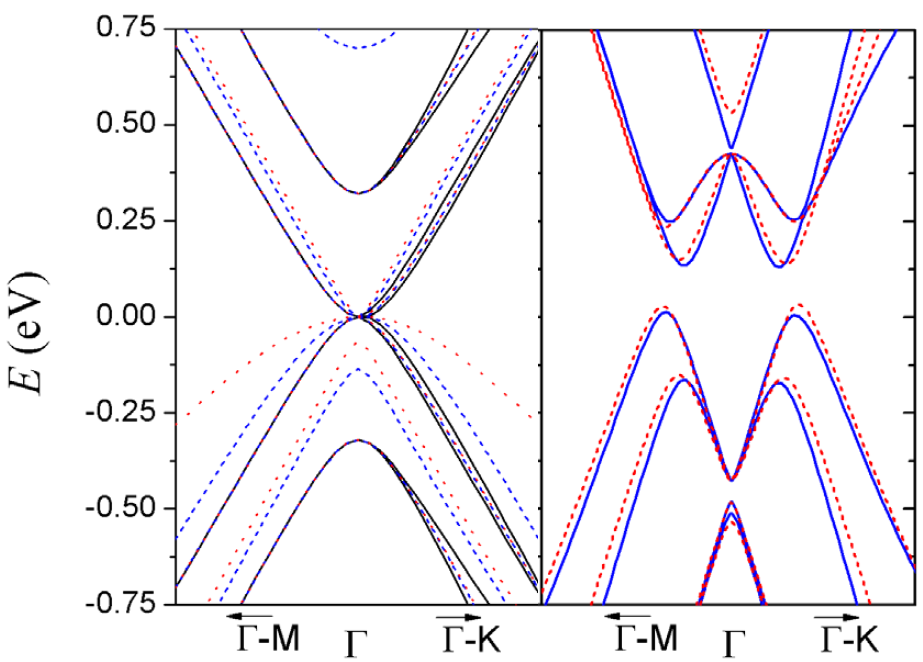}
\caption{\label{fig-tb-local} ({\it left panel}) Band structure for pure BLG (black full line) and for two short-range models for boron impurities in a $3\times 3$ cell, using an onsite potential of 4.2 eV at the impurity site and nearest-neighbor hoppings of -1.68 eV (red dotted line) and -3.12 eV (blue dashed line) between boron and carbon. ({\it right panel}) DFT (blue full line) and tight-binding (red dashed line) band structures of substitutional boron-doped BLG in a $3\times 3$ supercell. The tight-binding model includes long-range and short-range contributions of the impurity doping, as described in the text.}
\end{figure}

After an extensive search, we find that a minimum tight-binding model that reproduces the DFT band structures for ordered arrays of substitutional impurities is composed of a short-range and a long-range part. The short-range part is described by a change in the onsite matrix elements at the impuritity sites. The long-range part is modeled by a homogeneous-field-like $U$ onsite potential for all the atoms belonging to one of the layers and it arises from the difference in screened electrostatic potential in the two layers caused by the presence of the impurities in just one of them. Since our system is composed of a distribution of screened Coulomb impurities, for interdefect distances smaller than the Thomas-Fermi screening length in BLG \cite{hwang08}, as it is the case for the range of concentrations described in this work, a homogeneous field is sufficient to describe this contribution. It is important to note that the long-range contribution is fundamental to open a gap in BLG. We have tested several models containing only short-range modifications in the BLG hamiltonian, such as onsite-only or onsite plus nearest-neighbor-hopping terms. Such short-range models fail to open a gap in BLG. As an example, we show in the left panel of Figure \ref{fig-tb-local} the band structure for the $3\times3$ structure using two of those models (the pure BLG case is also shown for comparison), in which an onsite potential of $\epsilon_B$ = 4.2 eV is used at the impurity site to model boron and hopping parameters between the impurity and nearest-neighbor carbon atoms are varied between -30\% and +30\% from the C-C hopping value, which are rather extreme variations. As shown, a gap is not opened. The same result is obtained by varying the interlayer hopping between boron and carbon and by considering cells with different periodicities. \footnote {Strictly local model tight-binding potentials for substitutional impurities fail to open a gap in BLG because the zero-energy degenerate states at the Dirac point have zero amplitude at the defect site.}

The right panel of Figure \ref{fig-tb-local} shows the DFT and TB band structure of an ordered $3\times3$ array of
substitutional boron impurities, but the TB model now includes both short-range and long-range contributions of the impurity doping. The DFT (blue) is the same band structure shown in Figure ~\ref{fig-3x3}. The parameters $\epsilon_B$ = 4.2 eV and $U=0.852$ eV are used to fit the TB bands to the DFT ones. It is not necessary to vary the hopping elements in this case, as shown by the excellent agreement between the two curves, as compared with the poor agreement in the homogeneous-field model case (Figure \ref{fig-3x3}). This shows that the TB model with long-and short-range contributions provides a good description of the effects of substitutional impurities in graphene and therefore it can be used to explore disordered arrangements of such impurities.

\begin{figure}[h]
\centering
\includegraphics[width=8.7cm]{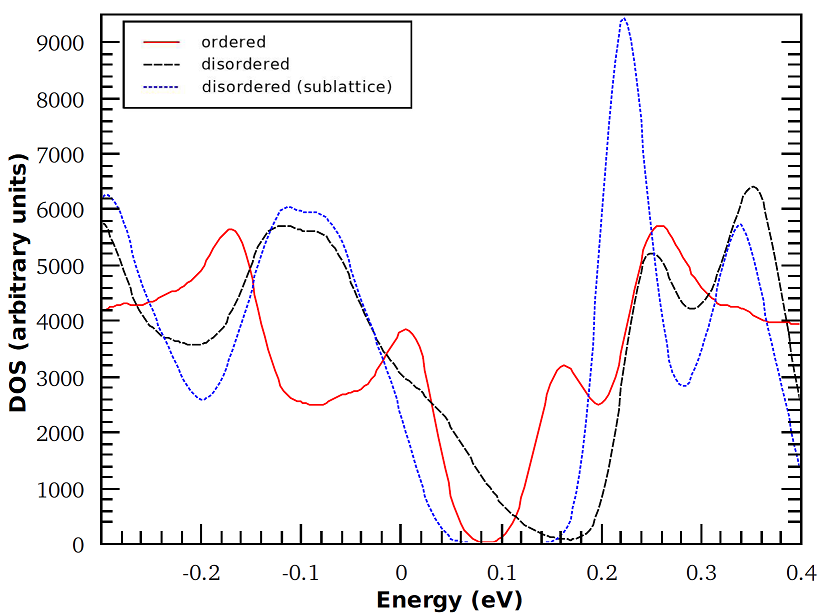}
\caption{\label{fig-dos} DOS near the Fermi level for the ordered $3\times 3$ structure (full red line) and for disordered structures within A sublattice (dotted blue line) and within both sublattices (dashed black line). DOS for disordered structures involve configurational averages for 20 realizations each. In all figures, discrete levels are Gaussian-broadened using a 0.02 eV width.}
\end{figure}

To achieve this goal, we use supercells with 10,000 sites and we distribute 278 boron atoms in one of the layers, so as to reproduce the same concentration as the ordered $3\times 3$ structure. Statistical averaging is performed over 20 realizations. In order to probe the effects of diferent sublattice distributions, we consider two different arrangements: (1) The boron atoms are randomly distributed over one of the sublattices only or (2)  the boron atoms are randomly distributed over both sublattices. The arrangement (1) is highly hyphotetical, but it is rather instructive to consider it, as we shall see. Surprisingly, the two choices lead to substantially different results for the electronic structure, as shown in Figure \ref{fig-dos}, which displays the density-of-states (DOS) near the Fermi level for the disordered arrangements (1) and (2), and for the ordered $3\times 3$ arrangement as well. Notice that the energy gap of 0.1 eV for the ordered structure (red full line), consistent with band structure of Figure \ref{fig-tb-local}. If a disordered arrangement is imposed to only one of the sublattices (blue dotted line) the gap opens up considerably, reaching almost 0.2 eV. However, and surprisingly, when disorder is imposed to both sublattices (black dashed line) the gap shrinks again and shifts to higher energy. This situation is more closely related to experimental realizations of such doped systems, in which sublattice control of impurity positioning is unlikely to be achieved. The intrincate dependence of the gap on different types of sublattice disorder in the substitutional case in BLG reveals the connection between sublattice symmetry and the band structure of graphene systems, as we recall that breaking sublattice symmetry is a way to open a gap in {\it monolayer} graphene.     
  
\section{\label{conclusion}Conclusions}
We described the effects of asymmetric doping in the band gap of BLG. Adsorbed potassium donors produce very similar band gaps, as a function of defect concentration, as an external and homogeneous electric field with varying magnitude. However, the behavior of substitutional impurities is quite different, leading to much smaller gaps and an acute sensitivity on the disorder and sublattice arrangement of impurities. We recall that, in a paradigmatic and successful model for donor and acceptor impurities in semiconductors, based on the envelope function and effective mass approximations, the short-range details of the impurity potential are quite irrelevant. So, one could expect that all donors or acceptors (both substitutional and adsorbed) should behave in a similar way. Therefore, it is certainly not obvious that this is not the case in graphene and we trace the different behaviors to the local symmetry of both kinds of impurities. Therefore, we conclude that models based on homogeneous fields can describe quite reasonably the gap opening induced by adsorbed donors, but care must be taken when the doping is related to substitutional foreign species.  

\begin{acknowledgments}
We acknowledge the financial support from the Brazilian funding agencies: CAPES, CNPq, FAPERJ and INCT-Nanomateriais de Carbono.
\end{acknowledgments}

\bibliography{Artigo}

\end{document}